\documentclass[pre,twocolumn,floatfix,superscriptaddress,longbibliography]{revtex4-2}
\usepackage{amssymb,amsmath,amstext}
\usepackage{graphicx}
\usepackage{subfigure}
\usepackage{amsthm}
\usepackage{verbatim}
\usepackage{dcolumn}
\usepackage{bm}
\usepackage{epsf}
\usepackage{xcolor}
\usepackage[colorlinks=true,citecolor=blue,linkcolor=blue,urlcolor=blue]{hyperref}
\newcommand{\ket}[1]{\left|#1\right\rangle}
\newcommand{\tr}[1]{\mathrm{tr}\left[#1\right]}
\newcommand{\bfm}{\boldsymbol}

\begin{document}
%===================================================================================================================
% Title & Abstract
%===================================================================================================================

\title{Approximate optimization, sampling and spin-glass droplets discovery \\ with tensor networks} 

\author{Marek M. Rams}
\email{marek.rams@uj.edu.pl}
\affiliation{Jagiellonian University, Institute of Theoretical Physics, \L{}ojasiewicza 11, 30-348 Krak\'ow, Poland}
\author{Masoud Mohseni}
\email{mohseni@google.com}
\affiliation{Google Quantum Artificial Intelligence Lab, Venice, California 90291, USA}
\author{Daniel Eppens} 
\affiliation{Google Quantum Artificial Intelligence Lab, Venice, California 90291, USA}
\author{Konrad Ja\l{}owiecki}
\address{Institute of Physics, University of Silesia, Uniwersytecka 4, 40-007 Katowice, Poland}
\author{Bart\l{}omiej Gardas}
\email{bartek.gardas@gmail.com}
\affiliation{Institute of Theoretical and Applied Informatics, Polish Academy of Sciences, Ba{\l}tycka 5, 44-100 Gliwice, Poland}
\affiliation{Jagiellonian University, Marian Smoluchowski Institute of Physics, \L{}ojasiewicza 11, 30-348 Krak\'ow, Poland}
\begin{abstract}
We devise a deterministic algorithm to efficiently sample high-quality solutions of certain spin-glass systems that encode hard optimization problems.
We employ tensor networks to represent the Gibbs distribution of all possible configurations.
Using approximate tensor-network contractions, we are able to efficiently map the low-energy spectrum of some quasi-two-dimensional Hamiltonians.
We exploit the local nature of the problems to compute spin-glass droplets geometries, which provides a new form of compression of the low-energy spectrum.
It naturally extends to sampling, which otherwise, for exact contraction, is $\#$P-complete. 
In particular, for one of the hardest known problem-classes devised on chimera graphs known as deceptive cluster loops and for up to $2048$ spins, we find on the order of $10^{10}$ degenerate ground states in a single run of our algorithm, computing better solutions than have been reported on some hard instances.  Our gradient-free approach could provide new insight into the structure of disordered spin-glass complexes, with ramifications both for machine learning and noisy intermediate-scale quantum devices.
\end{abstract}

\maketitle

\section{Introduction}

One of the most fundamental challenges for developing sufficiently advanced technologies is our ability to solve hard discrete optimization problems. 
These combinatorial problems have numerous applications across scientific disciplines and industries, in particular, machine learning and operations research. 
In the worst-case scenario, these problems require searching over an exponentially large space of possible configurations~\cite{Moore11}.

A general probabilistic, physics-inspired heuristics to sample the solution space of such problems is given by Markov chain Monte Carlo (MCMC) that relies on local thermal fluctuations enforced by Metropolis-Hastings updates~\cite{Metropolis49, Hastings70}.
This class includes simulated annealing~\cite{Kirkpatrick671} and parallel tempering (PT) algorithms~\cite{Hukushima_PT_2016,PTreview}. 
More advanced techniques combine specific probabilistic cluster-update strategies over a backbone algorithm from the MCMC family.
Those include Swendsen-Wang-Wolf cluster updates~\cite{Swendsen_Wang87, Wolf89}, Hodayer moves~\cite{Houdayer2001}, or Hamze-Freitas-Selbey algorithm~\cite{HF04, Selby14, Hen_2017}. 
However, these approaches either break down for frustrated systems~\cite{Wolf89}, or percolate above two-dimensions~\cite{Houdayer2001}, or assume random tree-like subgraphs~\cite{HF04, Selby14, Hen_2017}
that are not necessarily related to the actual structure of the low-energy excitations of the underlying problem. 

Another class of probabilistic physics-based approaches relies on quantum fluctuations to induce cluster updates. Those include adiabatic quantum computation~\cite{Nishimori_QA_1998, Lidar18}, dissipative quantum tunneling~\cite{Boixo16}, or coherent many-body delocalization effects~\cite{Kechedzhi18}. However, the potential computational power of such quantum processors is yet not well understood for noisy intermediate-scale quantum devices~\cite{Mohseni17, Preskill_NISQ_18}, as they could suffer from decoherence effects, finite control precision, or sparse connectivity graphs.

Correlations induce geometry in the state space. One of the key questions that lead to our work was whether it is possible to capture the underlying geometry of the combinatorial optimization problem with tensor networks~\cite{Murg07, Schollwock_10, orus14, Haegeman16, biamonte_review, Lewenstein18, orus2019tensor}.  Indeed, tensor networks receive the most attention in the context of weakly-entangled quantum many-body states~\cite{Eisert_review_entanglement} where they provide efficient and tractable decomposition allowing for successful digital simulations. For classical systems, among others, tensor-network contractions can be applied to compute the exact solution of specific optimization problems such as counting~\cite{Biamonte2015, Fast18}. An exact contraction of a generic tensor network is, however, a $\#$P-complete~\cite{Schuch_PEPS_complexity, gluza_peps_2018} task. In this article, we demonstrate a deterministic heuristic algorithm to systematically learn the low-energy spectrum of low-dimensional spin-glass complexes employing approximate tensor network contractions. In particular, we combine the latter with a branch and bound search strategy, where efficient utilization of the locality of interaction allows for a compressed description of the low-energy manifold based on a hierarchical structure of spin-glass droplets (excitations).

\begin{figure*}[t]
\begin{center}
  \includegraphics[width=1.99\columnwidth]{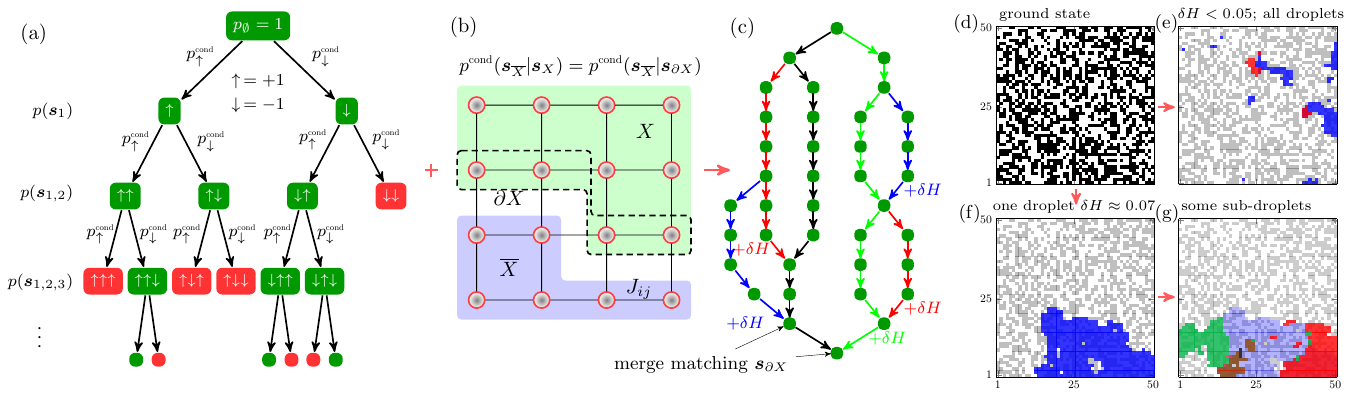}
\end{center}
  \caption{{Droplet revealing branch and bound strategy.}
  (a)~A tree search to find the most probable spin configurations for the Ising model~\eqref{eq:HcIsing}. At each depth of the tree, up to $M$ most probable configurations (marked green) are stored---here, we show $M=3$ for clarity. Marginal probabilities of the first $k$ spins, $p(s_1,s_2,\ldots,s_k)$, are calculated by approximate contraction of PEPS tensor network, see Fig.~\ref{fig:all}. (b)~Ising model with local interactions.
Conditional probability for spins in the region $\overline X$ (blue), conditioned on the given configuration in the region $X$ (green), depends only on the values of spins at the border $\partial X$. It is used in panel (c) to merge partial configurations with the same spins at an instantaneous border $\partial X$ (between spins which were, and were not, considered at a given level of the tree search), and reveal the structure of the spin-glass droplets. Here, black arrows depict the most probable path revealing the ground state, and other colors illustrate local low-energy excitations. In panels (d--g) we show an example of a single instance defined on square lattice $50\times50$ with random, uniformly distributed $J_{ij} \in [-1,1]$ (and weak local field $J_{ii} \in [-0.1, 0.1]$). A configuration with the lowest energy is in panel (d) [black dots represent $s_i=+1$]. In panel (e), we mark all clusters of spins flipping of which increases energy by $<0.05$. Disconnected---on the interactions graph---droplets can be flipped independently (we distinguish overlapping ones with different colors: blue and red).  In panel (f), we show a single, particularly large droplet connecting the ground state with a distant low-energy basin of attraction. Finally, in panel (g), we show a glimpse of the hierarchical structure of droplets: In red, green and black (distinguishing overlapping ones), we plot some clusters of spins which can be flipped following the flipping of the blue one in panel (f) [the latter is marked in panel (g) in light blue]. \label{fig:tree}}
\end{figure*}

Motivated by the topology of near-term quantum annealers, we consider the Ising Hamiltonian~\cite{Lucas14},
\begin{equation}
H(\bfm s) =  \sum_{\langle i, j\rangle \in \mathcal{E}} J_{ij} s_i s_j + \sum_{i =1}^N J_{ii} s_i,
\label{eq:HcIsing}
\end{equation}
where the couplings $J_{ij} \in \mathbb{R}$ are the input parameters of a given problem instance, with $N$ variables taking values $s_i=\pm 1$.  Here, we assume that the edges $\mathcal{E}$ form a quasi-two-dimensional structure, allowing us to try in that context established tensor-network contractions strategies.
In particular, we focus on the chimera graph, see Fig.~\ref{fig:all}(d), 
which is being realized in some quantum annealing processors~\cite{Lanting14}.

In this work, we represent the probability distribution $p(\bfm s) \sim \exp[-\beta H(\bfm s)]$ as tensor network equivalent to projected entangled pair states (PEPS)~\cite{nishino_peps, PEPS}. Approximately contracting the network allows one to efficiently calculate the probability of any configuration, including the marginal ones: 
\begin{equation}
 p{(s_1, s_2, \ldots, s_k)} \sim \tr{\mathcal{P}_{(s_1, s_2, \ldots, s_k)} e^{-\beta H(\bfm s)} },
 \label{pdo}
\end{equation}
with $\mathcal{P}_{(s_1, s_2, \ldots, s_k)}$ being a projector onto the subspace with a given configuration $(s_1, s_2, \ldots, s_k)$.  
Such approximate tensor-network contraction can be understood in entirely classical terms as an efficient method to construct and manipulate low-rank matrices to approximate the evaluation
of partition function or marginal probabilities.  In this context, message passing or belief propagation algorithms~\cite{BP03,SP2016} can be understood as some form of tensor network contractions that are exact over trees. 

In the rest of the article, we first discuss,  in Sec.~\ref{sec:II}, the branch and bound search that we use. We comment on the construction of the tensor network for the classical Boltzmann distribution, in particular for chimera graph, and its efficient contraction for conditional probabilities in Sec.~\ref{sec:III}, and collect exemplary results and benchmarks of our approach in Sec.~\ref{sec:IV}, followed by concluding remarks in Sec.~\ref{sec:V}. We provide additional details regarding the generation of one of the problem classes we use, the evidence on the conditioning of tensor network contraction, and its preconditioning in the Appendix.

\section{Droplet revealing branch and bound search}
\label{sec:II}

To extract the low-energy states from among exponentially many spin configurations, we employ branch and bound strategy, see Fig.~\ref{fig:tree}. In particular,  $k$th level of a binary tree in Fig.~\ref{fig:tree}(a)
contains partial states $(s_1, s_2, \ldots, s_k)$ together with their corresponding marginal probabilities. 
We explore the tree structure layer by layer, keeping at most $M$ partial configurations at a given step. To that end, at each depth, we branch $M$ current configurations into $2M$ new ones, taking into account one more spin (or $2^l M$, if we consider a group of $l$ new spins in one step). We then keep only those with the largest marginal probabilities, 
\begin{equation}
\label{cp}
\begin{split}
p{(s_1, s_2, \ldots, s_k, s_{k+1})} &= p{(s_1, s_2, \ldots, s_k)} \\
&\times p^{\scalebox{0.7}{cond}}(s_{k+1} | s_1, s_2, \ldots, s_k),
\end{split}
\end{equation}
with the last term being the conditional probability. A useful strategy to determine $M$ (making it step-dependent), is introducing a probability fluctuation cutoff $\delta_P$. In that case, we keep all the configurations among the considered ones, whose marginal probability divided by the maximal probability is larger than $\delta_P$.

More importantly, it is possible to leverage the locality of the problem, at the same time revealing the underlying geometries of the low-energy manifold. Indeed, for a configuration $\bfm s_X$ in the region $X = {(1, 2, \ldots, k)}$, the conditional probability in Eq.~\eqref{cp} depends only on the subconfiguration on the border $\partial X$, that consist of all spins in $X$ directly interacting with the region $\overline X = (k+1, k+2, \ldots, N)$. This idea is depicted in Fig.~\ref{fig:tree}(b), using a square lattice as an example.
Consequently, if two different configurations $\bfm s^1_X$ and $\bfm s^2_X$ coincide on the border $\partial X$, we can merge them in the tree search as depicted in Fig.~\ref{fig:tree}(c). 
This is evident from the chain rule in Eq.~\eqref{cp}, and the fact that $p(\bfm s_{\overline X} | \bfm s^1_X ) = p(\bfm s_{\overline X} | \bfm s^1_{\partial X} ) = p(\bfm s_{\overline X} | \bfm s^2_{\partial X} ) = p(\bfm s_{\overline X} | \bfm s^2_{X} )$. 
We seek for such configurations at each level of the tree search after branching and before discarding the improbable ones. On the one hand, this allows one to avoid repeating the calculation of the same conditional probabilities in Eq.~\eqref{pdo}, using only one branch (out of $M$) per a unique boundary configuration.

On the other hand, the more probable configuration of the two merged ones can be considered as the main branch. The other one, with larger or equal internal energy, defines a low-energy local excitation above the main branch, i.e., a spin-glass droplet. This excitation is naturally captured by the difference in spin orientations between $\bfm s^1_X$ and $\bfm s^2_X$. Subsequent merges result in a complicated structure consisting of both independent and nested excitations, as pictorially depicted in Fig.~\ref{fig:tree}(c). We keep track of only those up to some total excitation energy above the ground state.

Namely, to encode the low-energy spectrum, we associate a hierarchical structure (a tree) of droplet excitations above each branch in the branch and bound search. Such construction is not unique, which in our case has to do with the order of exploring the network, how the information about independent droplets is encoded, and how it is combined when the branches are merged. We discuss here two strategies that we employed in this work.

\begin{figure*} [t!]
\begin{center}
  \includegraphics[width=1.95\columnwidth]{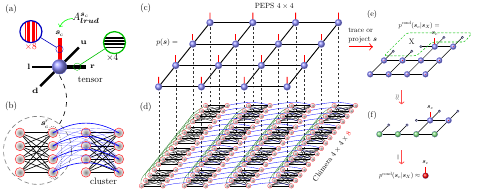}
\end{center}
  \caption{{Tensor network formalism to solve classical optimization problems on chimera-like graphs.}
  {(a,b)} A mapping explaining how PEPS tensors are assigned to groups of $l$ spins forming chimera-like graph. Each tensor has four virtual and one physical bond of sizes $D=d^{{\rm min}(m,n)}$ and $d^l$, respectively. Here, $d=2$ while $m$ is the number of spins in one group interacting with $n$ of those in the neighboring group.  For the chimera graph drawn in panel (d), $n=m=4$, where we indicate the interactions between the groups of $l=8$ spins with blue and green lines, while interactions within each group with black lines. 
 Adding more complicated interaction pattern shown with thin blue lines in panel (b) would not change $D$, see the main text. (c) The resulting tensor network allows one to represent the probability distribution $p(\bfm s) \sim \exp[-\beta H(\bfm s)]$ for the entire graph. (e) The conditional probabilities $p^{{cond}}({\bfm s_c} | {\bfm s}_X)$ are obtained by projecting the physical degrees of freedom in the region $X$ to given configuration ${\bfm s_X}$, and tracing out the remaining ones (marked with red dots).
 Black dots represent tensors completing the decomposition from adjacent spins in $X$  (see text), which amounts to selecting the sign of additional, effective local fields  acting on spins in blue tensors.
 Next, the approximate MPO-MPS scheme is invoked to collapse the network in a bottom-top fashion until only two rows remain. Finally, in panel (f), the remaining tensors can be exactly contracted to retrieve the desired conditional probabilities. \label{fig:all}}
\end{figure*}

We explore the effective two-dimensional (2D) network [see, e.g., Figs.~\ref{fig:tree}(b) and \ref{fig:all}(c) and \ref{fig:all}(d)] row after row, which sets a linear order for considered groups of spins, which is equivalent to the top-to-bottom order in the tree in Figs.~\ref{fig:tree}(a), and~\ref{fig:tree}(c). In the first approach, we use this order to decide which droplets can be flipped independently. Let us assume that we are merging two spin configurations, ${\bfm s}_X^1$ and ${\bfm s}_X^2$---and ${\bfm s}_X^1$ is becoming the main branch. The new excitation is defined by the spins where the two configurations differ, $e = {\bfm s}_X^1 \oplus {\bfm s}_X^2$ (with $\oplus$ referring to xor). Each $e$ has a beginning (the first node where the configurations differ) and an end (node where merging is happening) in the above-mentioned linear order. In the first approach, two excitations, $e_1$ and $e_2$, are considered independent if the end of $e_1$ appears before the beginning of $e_2$ (or the other way around). During merging of ${\bfm s}_X^2$ with ${\bfm s}_X^1$, $e$ is added to the list of excitations of ${\bfm s}_X^1$. All the excitations of ${\bfm s}_X^2$ independent of $e$ are discarded, as they should already be appearing as excitations of ${\bfm s}_X^1$. All excitations which are not independent (together with their subexcitations) are stored as sub-excitations of $e$. Employing such a procedure during the whole branch and bound strategy results in a structure similar to Fig.~\ref{fig:tree}(c). Finally, we can recover any low-energy configuration via backtracking -- they correspond to all the possible paths which lead from the last node back to the root. This procedure provides a one-to-one mapping of the low-energy spectrum of the problem of interest. Provided, of course,  that all the low-energy states were properly identified during the search (contractions of the tensor network were precise enough, the number of branches $M$ was large enough, etc.).

In the second approach, excitations $e_1$ and $e_2$ are considered independent if no spin in $e_1$ is connected (via some nonzero $J_{ij}$) with some spin in $e_2$. During merging of ${\bfm s}_X^1$ and ${\bfm s}_X^2$, we discard the excitation $e$ if it is not singly connected on the graph of $J_{ij}$'s. Otherwise, it is added to the list of excitations of ${\bfm s}_X^1$. Excitations of ${\bfm s}_X^2$ which are not independent of $e$ are added as its sub-excitations.
In that case, for instances with discreet $J_{ij}$ and degenerate spectrum, adding in this procedure a small random noise to $J_{ij}$ (lifting degeneracies) allows to resolve some possible ambiguities---as a droplet which is not singly-connected is going to have larger energy than each of its singly-connected parts. Finally, the low-energy spectrum is recovered by considering all sets of droplets (together with sub-droplets of flipped droplets) that can be flipped independently. This strategy, however, does not correctly solve all possible ambiguities when two partial configurations with many layers of hierarchical excitation structure are merged. Consequently, it is not giving a full one-to-one mapping of the low-energy spectrum. 

Other strategies are possible, and we leave exploring them as future work. At the same time, in both discussed approaches, one may introduce a threshold and discard small-size excitations $e$ that flip too few spins, below some cutoff. Introducing such a threshold allows one to get compact course-grained information about the low-energy spectrum of the instance of interest.

It is worth stressing that the droplets that we found here are consistent with the droplet picture for the Edwards-Anderson model of spin-glasses~\cite{Fisher86, Kawashima99}. In particular, in Figs.~\ref{fig:tree}(d)--\ref{fig:tree}(g) we show an example of a single random instance defined on a square lattice with nearest-neighbor interactions. Therein, we show a snapshot of an identified hierarchy of droplets, i.e., groups of spins flipping of which switches between particular low-energy configurations. 

By iterating such a branch and bound procedure down to the last site, we produce a candidate for the ground state, structure of low-energy states build on top of it, as well as the largest marginal probability that was discarded in the process, $p_d$. In principle, this could allow one to verify if the ground state indeed has been found.  As $p_d$ bounds probabilities of all configurations which have been discarded,  the maximal calculated probability (corresponding to the state with the lowest energy) being greater than $p_d$ would be a sufficient condition for such verification---assuming we had an oracle to precisely calculate the partition functions. In practice, as we employ approximate tensor network contractions, this remains heuristic evidence.

\section{PEPS tensor network for conditional probabilities}
\label{sec:III}

To execute the outlined algorithm one needs to effectively calculate conditional probabilities $p^{\scalebox{0.7}{cond}}(s_{k+1} | s_1, s_2, \ldots, s_k)$ (more generally, probabilities for a group of $l$ spins ${\bfm s_c}$) to employ the chain rule in Eq.~\eqref{cp}. The idea is to simultaneously encapsulate all of them by a two-dimensional PEPS tensor network. Finding an approximate PEPS representation of the ground state or the thermal state of a 2D {\it quantum} system is a challenging problem and typically requires iterative variational optimization, see, e.g., Refs.~\cite{orus14, Vanderstraeten16, Corboz_2017, Czarnik_2015}. However, for a classical spin system such as in Eq.~\eqref{eq:HcIsing}, the construction of a thermal state is exact and identical to that of its partition function~\cite{nishino_peps, PEPS}.

Indeed, consider two sites, say $i$ and $j$, connected by an edge with $J_{ij}$. A natural decomposition which one can explore reads, 
\begin{equation}
e^{-\beta J_{ij} s_i s_j} =
\sum_{\gamma=\pm 1} B^{s_i}_{\gamma} C^{s_j}_{\gamma},
\label{asvd}
\end{equation}
with $B_{\gamma}^{s_i}=\delta_{s_i \gamma}$, and $C_{\gamma}^{s_j}=e^{-\beta J_{ij} \gamma s_j}$ ($\delta_{kl}$ is the Kronecker delta). These tensors serve as basic building blocks for all our constructions. Albeit not unique, Eq.~\eqref{asvd} has the advantage of containing only nonnegative terms increasing numerical stability. Even this property, however, does not ensure uniqueness, and we further explore this in the preconditioning procedure, as described in the Appendix.

Here, we focus on the chimera graph depicted in Fig.~\ref{fig:all}(d). The building block of this graph consists of a group of $l=8$ spins. Only $4$ spins in a given cluster interact with those in the neighboring cluster. We explore such a grouping of spins, and with each group of $8$ spins, we associate a tensor
\begin{equation}
\label{cpeps}
A^{\bfm s_c}_{\bfm{lrud}} 
= e^{-\beta H(\bfm s_c)}
B_{\bfm l}^{\bfm s^l_{c}}
C_{\bfm r}^{\bfm s^r_{c}}
B_{\bfm u}^{\bfm s^u_{c}}
C_{\bfm d}^{\bfm s^d_{c}}.
\end{equation}
Here, ${\bfm s_c}$ collects the spins in the considered cluster and $\bfm s^l_{c}, \bfm s^r_{c}, \ldots$ are the subsets of those spins which are interacting with the neighboring clusters to, respectively, left, right, etc. The interactions with the cluster to the left are encoded as $B_{\bfm l}^{\bfm s^{\bfm l}_{c}} = \prod_{k=1}^4 B^{s^{l_k}_c}_{l_k}$, where $\bfm l = (l_1, l_2, l_3, l_4)$ collects the virtual indices [$\gamma$ in Eq.~\eqref{asvd}] for respective decompositions. The same holds for the remaining directions. 
As a result, each PEPS tensor has now one physical index $\bfm s_{i}$
of size $2^8$ and $4$ virtual ones: $\bfm l, \bfm r, \bfm u, \bfm d$
-- each of size $D = 2^4$. 
Finally, $H(\bfm s_{c})$ is the inner energy of the group---where the sum in Eq.~\eqref{eq:HcIsing} is limited to the subgraph formed by spins $\bfm s_c$.  Finally, combining all the tensors leads to a representation of the probability distribution as
\begin{equation}
\exp[-\beta H(\bfm s)] \sim \sum_{\{ \bfm{k} \}} \prod_{c^i} A^{\bfm s_{c^i}}_{{\bfm{ {k^i} } }}, 
\end{equation}
where $c^i$ numerates all the clusters, and the sum (or effectively tensor contractions) is performed over all the repeated virtual indices connecting the neighboring clusters, see Figs.~\ref{fig:all}(a) and Fig.~\ref{fig:all}(c). In practice, for calculation of the partition function or marginal probabilities one first traces out physical degrees of freedom $\bfm s$.

The above construction is sufficient to build the PEPS representation of the chimera graph.  However, it is worth noticing at this point that one can introduce a substantially more complicated interaction pattern between neighboring clusters. For example, in Fig.~\ref{fig:all}(b), we show an example where each of $4$ qubits couples not to one of its neighbors (as in the chimera geometry), but to all of them (depicted as transparent blue lines). One can still capture this pattern without enlarging the bond dimension $D$. Indeed, suppose that site $i$ talks to more than one of its nearest neighbors, say $j$ and $k$. Then there is only one bond that goes through this interaction, i.e.,
\begin{equation}
e^{-\beta J_{ij} s_i s_j - \beta J_{ik} s_i s_k} =
\sum_{\gamma=\pm1} B^{s_i}_{\gamma} \left(C^{s_j}_{\gamma} C^{s_k}_{\gamma}\right).
\label{asvd2}
\end{equation}
The same argument applies when there are $4$ sites involved. With this strategy, one can easily encode a variety of quasi-2D graphs. In particular, $m$-spins to $n$-spins interaction between neighboring clusters can be captured by a PEPS with the bond dimension $D=2^{\min(m,n)}$.  Interactions at a longer range, e.g., between the next-nearest clusters, are also possible to construct.
To that end, as a building block of PEPS tensors, one can use the matrix product operator (MPO) decomposition [see Eqs.~\eqref{gmpo} and \eqref{BC_definition} in the Appendix] that generalizes the two-site decomposition in Eq.~\eqref{asvd} to several sites. Such MPO's can overlap,  increasing the bond dimension of the resulting PEPS---multiplying bond dimensions of MPOs building the network. 

Finally, we can focus on the calculation of conditional probabilities $p^{\scalebox{0.7}{cond}}({\bfm s_c}| {\bfm s}_X)$ in Eq.~\eqref{cp}. To that end, we first project on a given configuration ${\bfm s}_X$ in the region $X=(1,2,\ldots,k)$, and trace out all the remaining  degrees of freedom apart from $\bfm s_c$. This results in the network in Fig.~\ref{fig:all}(e), where the black dots represent tensors $B$ or $C$, completing the decomposition in Eq.~\eqref{asvd}, projected on desired configuration (limited to the spins directly interacting with $s_{k}, s_{k+1}, \ldots, s_{N}$ in the lower half of the network).  Now,  the conditional probabilities follow from collapsing that network, see Figs.~\ref{fig:all}(e) and \ref{fig:all}(f).

While the tensor network representation in Fig.~\ref{fig:all}(e) is exact, extracting information from it is still a \#{}P task. Although there are approximate contraction schemes one can utilize~\cite{Lewenstein18}, 
it is not obvious {\it a priori} how well they will perform in practice, particularly for disordered systems considered here.
 In this article, we employ a matrix product state (MPS)--matrix product operator (MPO) based approach~\cite{Murg07}.
 The idea is depicted in Figs.~\ref{fig:all}(e) and \ref{fig:all}(f). Essentially, the first row of the grid shown in
 Fig.~\ref{fig:all}(e) can be treated as a vector in
 high-dimensional virtual space, which has a natural
 underlying tensor structure of MPS. Adding another row (viewed as
 MPO) enlarges this MPS representation. Therefore, to prevent its exponential growth when yet other rows are added, truncation of the bond dimension is necessary. It results in a series of boundary MPSs with limited bond dimensions $\chi$ [there is only one in Fig.~\ref{fig:all}(f), marked green, approximating two rows of blue tensors in Fig.~\ref{fig:all}(e)].
They are found sequentially by minimizing their distance from the enlarged previous ones. This distance quantifies an error of a single truncation (see Appendix). Finally, the network in Fig.~\ref{fig:all}(f) can be contracted (numerically) exactly, resulting in the sought-after conditional probability.

\begin{figure} [t!]
\begin{center}
  \includegraphics[width= \columnwidth]{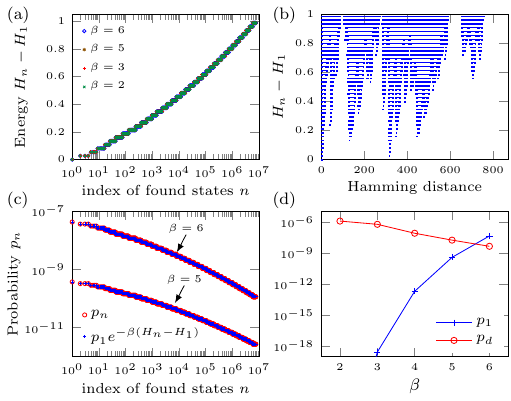}
\end{center}
  \caption{{Low-energy spectrum of the Ising Hamiltonian defined on the chimera topology.}
In panel (a), we show $\sim10^7$ different low-energy solutions for a hard structured problem of $N=2048$ variables---all of which were found in one run of the algorithm. 
We plot their Hamming distance from the ground state (i.e., the number of spins where the two configurations differ) in panel (b). It indicates that our method can sample solutions differing by $\sim O(N)$ spins.
In panel (c), we show the corresponding probabilities for numerically least stable large $\beta$'s.  In this case, we can see full consistency between the probabilities obtained from the contraction of the PEPS network and the Boltzmann weights calculated from configurations energies. Finally, in panel (d), we plot the probability of the ground state $p_{1}$ that we found together with the largest discarded probability $p_d$. Here, with increasing $\beta$, we were able to guarantee $p_d < p_{1}$, indicating that the ground state was not missed. Nevertheless, the same full low-energy spectrum is recovered for all values of $\beta$ in panel (a).
Instances were defined with discreet $J_{ij}$ with $dJ=\frac{1}{75}$, which results in visible discreetness of energies found. We focused here on one having many distinct local minima in panel (b) and used $M=2^{14}$.
}
   \label{fig:results}
\end{figure}

The outlined algorithm is deterministic, with the running time scaling polynomially with the control parameters.
The numerical cost of the preprocessing step, where the boundary MPSs are calculated, scales as $O(N D^4 \chi^3+N D^4 d^l)$.
Those are related to the truncation of boundary MPS and tracing PEPS tensors, respectively. Here $\chi$ is the maximal bond dimension of boundary MPS,
$D$ are the virtual bond dimensions of the PEPS tensor, and $d^l$ denotes its physical dimension, see Fig.~\ref{fig:all}.
The leading cost of calculating probabilities in the branch and bound search scales as $O(N M \chi^2 D^2 +N M D^2 d^l)$.
We should stress, nonetheless, that even in the ideal case of an oracle giving exact probabilities, certifying that the ground state has been found may require $M$ scaling exponentially with $N$.
At the same time, for ill-conditioned MPO, the error of the previous truncation can, in the worst-case, grow exponentially during the procedure.  Additionally, increasing the control parameter $\chi$ to obtain better accuracy may require more than the standard double numerical precision (used in this work), making thereof a limiting factor of the numerical simulations.

\section{Results and benchmarks}
\label{sec:IV}

Apart from instances with coupling drawn from independent distributions [as the one in Figs.~\ref{fig:tree}(d)--\ref{fig:tree}(g)], we have tested our algorithm with sets of instances that were specifically designed to be hard for classical heuristic approaches based on local updates.
In particular, we have used new {\it droplet instances} (see Appendix), which have many embedded skewed
droplets/clusters with a power-law distribution over various sizes up to a
length-scale of $O(N)$. It makes them hard for probabilistic heuristic algorithms that rely on local updates.  In Fig.~\ref{fig:results}, we show the results for a single instance consisting of a full set of low-energy states. While larger $\beta$ allows to ``zoom in" on low-energy states better, it also renders the tensor network contraction numerically ill-conditioned. Thus, one cannot provide tight bounds on the possible errors of calculated probabilities. Nevertheless, the method that we present here can provide empirical guarantees by verifying the consistency of the results obtained for different $\beta$'s and different ordering of contractions, see Appendix.
For intermediate $\beta=3$, setting the time limit per instance at half-an-hour (running on a single core and performing contraction from all four directions), we can find the ground states (i.e., the lowest
energies ever identified by us) for all 100 test instances. We provide times-to-solutions for some other reference solvers in Table~\ref{tab:tts}, where we focus on time to different approximation ratios (defined as the ratio between the excitation energy above the ground state energy and the total width of the whole energy spectrum). For reference,  the excitation energy $dH=1$ corresponds to the approximation ratio $\simeq 1.5 \times 10^{-4}$ for $N=2048$ in Fig.~\ref{fig:results}.

\begin{table}
\begin{tabular}{l|c|ccc}
 Method & approx. ratio & $N=512$ & $N=1152$ & $N=2048$ \\
\hline
\hline
This article & g.s. & 30s& 150s & 450s \\
\hline
PT (adaptive) & g.s. & 800s & --- & --- \\
\hline
PT (geometric)& $0.01$ & 0.53s & 4.16s & --- \\
PT (geometric)& $0.005$ & 2.51s & 56.4s & --- \\
PT (geometric)& $0.001$ & 158.4s & timed-out & --- \\
PT (geometric)& $0.0001$ & 897.6s & timed-out & --- \\
\hline
\hline
DWave 2000Q$_6$ & $0.01$ & $0.003$s & $0.006$s & $0.02$s \\
DWave 2000Q$_6$ & $0.005$ & $0.2$s & timed-out & timed-out \\
DWave 2000Q$_6$ & $0.001$ & timed-out & timed-out  & timed-out \\
\hline
\hline
\end{tabular} 
\caption{{Comparison of times to solutions for selected solvers.} We provide the median time from 100 droplet instances. The time for our approach is for a single run using a single-core with $\beta=3$, bond dimension $\chi=16$ and relative probability cutoff $\delta_P=10^{-3}$; parameters that are fully sufficient to reach the ground state within this metric. The times for PT algorithms are based on the number of MC sweeps, with $0.00005$s per MC sweep for $N=512$ and $0.00011$s for $N=1152$ (running on a single-core on the same machine). For probabilistic solvers, we estimate time to $99\%$ probability of success. Adaptive PT is for optimized hyper-parameters, including an adaptive profile of temperatures with $12$ replicas. For geometric PT, we have $25$ temperatures distributed geometrically between $\beta=$ 0.001 and 10. For quantum annealing, we run the experiments on a DWave 2000Q machine. Due to some inactive qubits in the machine, we have dropped them from the instances -- again using the approach of this paper to find the reference ground-state solutions for modified instances. For each instance, we run $2500$ repetitions ($1000$ for $N=512$), optimizing over annealing times $5$, $20$, and $200\mu$s. No overheads over pure annealing time are included.  Time-out indicates that we have not been able to find a single solution of the desired quality within our experiment, i.e., reaching a solution within a given approximation ratio.  \label{tab:tts}}
\end{table}

We have also tested the algorithm on the set of {\it deceptive cluster loops}~\cite{helmut_deceptive_2018} with parameter $\lambda=7$, for which they are expected to be the hardest for classical heuristics.
We recovered the reference the lowest energies, found with the help of other algorithms~\cite{helmut_deceptive_2018}, in $\sim97\%$ cases. In the remaining $\sim3\%$ we were
able to identify a state with better (lower) energy than the provided
referenced ones. Those instances offer a challenging test for our approach as they exhibit a humongous ground state degeneracy. We find its median to be $\sim10^{14}$ for $N=2048$.

\begin{figure} [t!]
\begin{center}
  \includegraphics[width= \columnwidth]{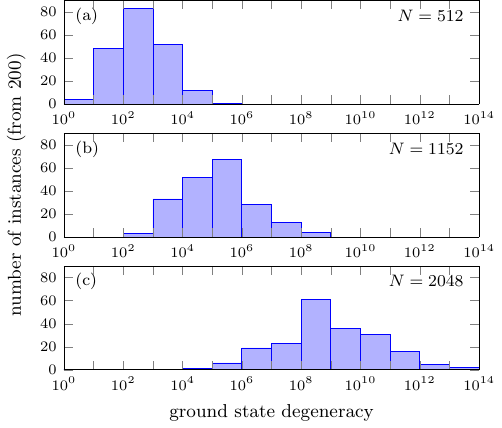}
\end{center}
  \caption{{Counting the ground state degeneracy.} Here, we consider the instances on the chimera graph drawn from a uniformly distributed with $J_{ij} \in \{\pm 1, \pm 2, \pm 4\}$ (without local fields, i.e., with $J_{ii}=0$). The same class of instances has been considered in the context of fair sampling of PT and PT+ICM algorithms in Ref.~\cite{zhu_fair_2019}. We show the histogram of degeneracies found by our approach, where we can go beyond $N=1152$ studied in Ref.~\cite{zhu_fair_2019}.
   \label{fig:degeneracy}}
\end{figure}

Finally, we benchmark our algorithm with regard to performing a fair sampling. To that end, we focus on instances with integer couplings and count the identified ground state degeneracy. We follow Ref.~\cite{zhu_fair_2019}, which studied fair-sampling properties of PT and PT+ICM (isoenergetic cluster moves~\cite{Zhu_PT+ICM_2015}; related to Houdayer's moves~\cite{Houdayer2001}) algorithms, while using instances with random, uniformly distributed $J_{ij} \in \{\pm 1, \pm 2, \pm 4\}$, including for the chimera graph geometry of up to $N=1152$ spins. We present our results for such instances in Fig.~\ref{fig:degeneracy}. Assuming that the ground state is found (we increase our confidence in that by running CPLEX solver \cite{cplex} for all instances, which have not reported any energies better than those found by our method), our approach directly counts the identified ground-state configurations. As such, strictly speaking, we provide a lower bound on the ground-state degeneracy. Utilizing the merging strategy outlined in Sec.~\ref{sec:II} allows to greatly enhance the efficiency of the process -- indeed, counting is performed here with no additional cost comparing to the identification of a ground state configuration. For smaller system sizes, our results seems to be consistent with the ones in Ref.~\cite{zhu_fair_2019} for the same distribution of couplings (We note that for $N=1152$ we observe some ground-state degeneracies approaching $10^8$, while only the numbers below $10^6$ have been previously reported).

\section{Conclusions}
\label{sec:V}

In summary, we demonstrated how tensor networks representation of
spin-glass problems could lead
to a profound insight into their low-energy landscape.
We performed approximate tensor contraction using an iterative MPS-MPO construction. One could explore alternative 
tensor contraction schemes based on renormalization group
techniques~\cite{Lewenstein18}. Also, the droplet finding algorithm introduced here can be
combined with Monte Carlo techniques to introduce nontrivial
nonlocal moves. We mainly focused on problems on chimera graphs that are currently being realized in quantum annealers~\cite{Wittek18}. 
It remains to be seen, however, how well our approach will perform for
the next generations of quantum annealers that utilize graph known as pegasus, which will have a higher degree of connectivity~\cite{dattani2019pegasus}. Answering  that question could strongly influence future hardware directions of quantum annealing. 

{\it Note added}: Recently, related work appeared where a closely related tensor-networks--based sampling strategy has been combined with the Metropolis-Hastings Markov chain acceptance rule to improve the stability of sampling from the thermal distribution~\cite{Frias_Collective_21}.

\begin{acknowledgments}
We thank D-Wave Systems Inc. for sharing the {\it deceptive cluster loops} instances, Jarrod McClean and Jacek Dziarmaga for careful reading of the manuscript, and Helmut Katzgraber and Andrew Ochoa for correspondence on the results of Ref.~\cite{zhu_fair_2019}.
This work was supported by National Science Center (NCN), Poland
under Projects No. 2016/20/S/ST2/00152 (B.G.), No. 2020/38/E/ST3/00269
(K.J.), and NCN together with European Union through QuantERA ERA NET
Program No. 2017/25/Z/ST2/03028 (M.M.R.). B.G also acknowledes support
of Foundation for Polish Science (FNP) under grant No.
POIR.04.04.00-00-17C1/18-00.
We acknowledge receiving Google Faculty Research Award 2017 (MMR) and 2018 (MMR and BG).
We make our implementation of the algorithm, together with the {\it droplet instances} and the instances we used to test counting of the ground state degeneracy, publicly available on GitHub~\cite{git}.
\end{acknowledgments}

\begin{appendix}
\section{}
\renewcommand{\theequation}{A\arabic{equation}}
\setcounter{equation}{0}

In this Appendix, we first describe the procedure for generating {\it droplet instances}---one of the instance classes we used to test our approach. Second, we discuss a complementary algorithm employing decomposition of the probability distribution as a matrix product state. It can be used to test---and at the same time better appreciate the performance of---the approach introduced in the main text. Third, we provide additional information regarding the contraction of the PEPS tensor network representing probability distribution for a quasi-two-dimensional lattice. General discussion of errors, as well as a heuristic (gauge) preconditioning that we use, are also included. 

\subsection{Generation of structured droplet instances on chimera graphs.}
We explored structured instances on the chimera graph (D-Wave quantum machine). The construction of these instances at the high level can
be understood with the following generator: 

\renewcommand{\labelenumii}{\Roman{enumii}}
\begin{enumerate}
\item Local fields: Draw $J_{ii}$ coefficients randomly from a probability density function (PDF) [e.g., flat or Gaussian] centered at zero with small standard deviation,
e.g., $0.1$. 
\item Background random spin glass: Draw nonclustered $J^r_{ij}$ from another PDF centered at zero with $\max |J_{ij}^r| > k_r\times\max|J_{ii}|$, where $k_r$ is a constant factor around $5$--$10$ such that they are much stronger that local fields.
\item Generate a power-law distribution of cluster sizes with $p(n_{\rm edges}) = n_{\rm edges}^{-\gamma}$, where $n_{\rm edges}$ is the number of edges forming a cluster, and an exponent $\gamma$ is such that $1 \leq \gamma \leq 3$.
\item Generating structured droplets:
Plant the seed of a droplet by drawing a random edge on the graph representing the problem instance and grow randomly connected clusters with size given by probability distribution $p(n_{\rm edges})$ over the graph topology of background random spin-glass system. Now for each edge in the cluster attribute a random $J^c_{ij}$  from a different PDF such that $\max |J_{ij}^c|
> k_c \times \max |J_{ij}^r|$ where $k_c$ is a constant factor between $2$ and $10$.
In other words, we boost certain connected edges from the background spin glass generated by first two steps by a factor $k_c$. The size of each connected cluster is given by $p(n_{\rm edges})$ and their shapes is completely random. 

\item Repeat the last step until a desired number of clusters are generated. The procedure needs to stop before the clusters percolate (in our instances only $5-10\%$ of overall background edges contribute to clusters)    
\end{enumerate}

This construction leads to instances that typically have many embedded
droplets with a power-law distribution over various sizes up to length-scale of $O(N)$. 
These instances are generally hard to solve for
probabilistic heuristic algorithms, such as simulated annealing, that
rely on local updates that are inefficient for flipping the underlying
clusters. 
Moreover, the droplets
typically have fractal geometry and thus are hard to be characterized 
by known cluster finding algorithms. 
The instances employed in this work had $J_{ij} \in [-5, 5]$ with the discrete step $dJ = \frac{1}{75}$.
We include them in the public repository in Ref.~\cite{git}.

\subsection{Matrix-product-states based approach}
\label{sec:mps}
The PEPS tensor network discussed in the main text incorporates a quasi-2D structure of chimera-like graphs. It is crucial when dealing with large problems where $N\sim 10^3$. However, any system, in particular 2D, can be considered as a 1D chain. One can explore this further to build another representation for the probability distribution in Eq.~\eqref{pdo} of the main text. A different algorithm can then be devised to benchmark against the PEPS approach for smaller systems, $N\sim 10^2$. This method is based on matrix product states (MPS) and their properties~\cite{Vidal03, Murg07, Schollwock_10}. Closely related, matrix product states representations were considered in the context of (nonequilibrium) classical stochastic processes \cite{Temme_stochastic_MPS, Jaksch_classical_MPS}, counting \cite{Chamon2012}, or more recently machine learning \cite{Stoudenmire2016,Han2018}.
We also use it to briefly introduce the main techniques of the MPS toolbox, which are used in the main text to contract the PEPS network via the boundary-MPS approach.

\subsubsection{Basic concepts}

Searching the probability rather than energy space is closely related to the paradigm of quantum computation.
To better understand why that is the case, we transform the classical Ising Hamiltonian as defined in Eq.~\eqref{eq:HcIsing} of the main text 
onto its quantum counterpart, $\mathcal{H}=H( \bfm {\hat \sigma}^z)$. Now, $\bfm {\hat \sigma}^z=(\hat \sigma^z_1, \dots, \hat \sigma^z_N)$ 
denote  Pauli operators acting on a local space $\mathbb{R}^2$. Obliviously, any classical solution 
$(s_1, \dots, s_N)$ translates naturally onto an eigenstate of $\mathcal{H}$ and \emph{vice verse}.

From a mathematical viewpoint, the Hamiltonian $\mathcal{H}$ does not simplify the original problem. It does, nonetheless, 
points to a possible strategy that could be utilized by classical computers to find $m \ll d^N$ lowest energy states. According to the Gibbs 
distribution, $\mathcal{\rho} \sim \exp(-\beta \mathcal{H})$, these states are also the most probable ones at a given temperature $1/\beta $. 
Therefore, one could prepare a quantum system in a superposition of all possible configurations, $|\bfm s\rangle = |s_1, s_2, \ldots s_N \rangle$, that is to say:
\begin{equation}
\label{rho}
\ket{\rho} \sim \sum_{\bfm s} e^{-\beta \mathcal{H}/2}\ket{\bfm s}.
\end{equation}
One could then perform a measurement, which for all intents and purposes is treated here as a black box~\cite{wheele83}. As assured by the laws of quantum mechanics, the low-energy states would be the most probable outcomes of such an experiment.
There are two paradigms involved in this scenario. First, one has to do with how all possible combinations are stored efficiently via a quantum superposition. Second is the information extraction via a suitable measurement that ultimately leads to the desired outcome. 

Similarly, the algorithm of the main text has two essential steps. First, we encode the probability distribution of all classical configurations as the PEPS tensor network. 
As we argued, the latter network provides a natural representation for such distribution. The extensive use of contraction techniques enables one to approximately calculate any marginal probability.  Note, as the network collapses, the information spreads across the entire system. It results in a highly nontrivial update that other heuristic methods lack. Next, we extract the desired number of states with the largest probability amplitudes.  In this analogy, instead of performing a quantum measurement, we search a probability tree, see Fig.~\ref{fig:tree} in the main text (which, however, allows us to additionally obtain compressed information about the structure of the low-energy excitations when we utilize locality of interactions).

Another approach, which can be naturally tested, is to approximately represented state $|\rho\rangle$ as a MPS,
\begin{equation}
\label{eq:mps}
\ket{\rho}  \approx  \sum_{\bfm s} M^{s_1} M^{s_2} \cdots M^{s_N}\ket{\bfm s}.
\end{equation}
Here each $M^{s_n}$ is a matrix, maximally of size $\chi \times \chi$, where the bond dimension $\chi$ controls the quality of such approximation.

\begin{figure} [t!]
    \begin{center}
        \includegraphics[width=0.9\columnwidth]{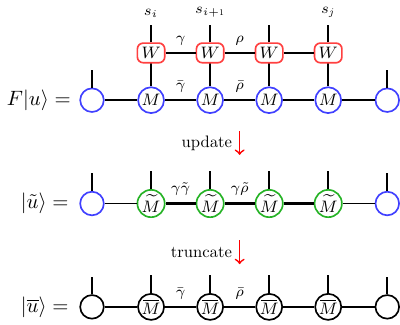}
    \end{center}
    \caption{{MPO-MPS scheme}. First, a MPO representing a gate $F= G_{i, i+1} G_{i, i+2}\ldots G_{i,j}$ [Eqs.~\eqref{eq:Fgate}--\eqref{BC_definition}] is applied to appropriate sites of state $|u\rangle$, see Eq.~\eqref{mpsMpo}. This results in increase of the bond dimension, here
        by a factor of $D=2$ (the bond dimension of MPO). Consequently, a truncation scheme is employed to approximate MPS to maintain its bond dimension at a manageable size. 
    }
    \label{fig:mpo_mps}
\end{figure}

To obtained the desired MPS, we begin with the Hadamard state $\sum_{\bfm s}\ket{\bfm s}$, i.e., an equal weight combination of all possible classical states, that has a trivial MPS representation $M^{s_n = +1} = M^{s_n = -1} = 1$ with bond dimension $\chi=1$. 
The MPS approximation in Eq.~\eqref{eq:mps} can be obtained by sequentially applying both the two-site gates, 
\begin{equation}
G_{i,j}=\exp(-\tau J_{ij} \sigma^z_{i} \sigma^z_{j}),
\label{eq:Fgate}
\end{equation}
acting on the edge $(i,j)$, and one-site gates $e^{-\tau J_{ii} \sigma^z_i}$. 
In the context of MPS it is convenient to simultaneously consider action of all gates sharing a common site $i$, i.e., 
\begin{equation}
F = {G}_{i,j_1} {G}_{i,j_2} ... {G}_{i,j_L}.
\label{eq:Ogeneral}
\end{equation}
One may represent it as matrix product operator (MPO) with bond dimension $2$. Namely,
\begin{equation}
\label{gmpo}
F = \sum_{\bfm s, \bfm {s'}} W^{s_i s_i'}  W^{s_{i+1} s_{i+1}'} \cdots  W^{s_{j_L} s_{j_L}'} | \bfm s \rangle \langle \bfm {s'} |,
\end{equation}
where for simplicity we assume that $i < j_1 < \ldots < j_L$.
For the classical partition function, all $W$'s are diagonal in physical indices $W^{s_m s_m'}= W^{s_m s_m} \delta_{s_m, s_m'}$.  
We then have $W^{s_i s_i}_{\gamma} =  B^{s_i}_{\gamma}$ at site $i$, and $W^{s_m s_m}_{\gamma \gamma'} =  C^{s_m}_{\gamma} \delta_{\gamma \gamma'}$ for $m =i+1, i+2, \ldots, j_{N}$. The basic building blocks read
\begin{eqnarray}
B_{\gamma}^{s_i} &=&  \delta_{s_i \gamma},   \nonumber \\
C_{\gamma}^{s_j} &=& e^{-\tau \gamma J_{ij} s_j},  \label{BC_definition}
\end{eqnarray}
with the virtual index taking values $\gamma = \pm 1$.
Finally, one-site gate $e^{-\tau J_{ii} \sigma^z_i}$ acts trivially at site $i$. Thus, it can be easily incorporated into Eq.~\eqref{eq:Ogeneral} by rescaling $B_{\gamma}^{s_i}$ by such factor.
Note that with such construction, all coefficients appearing in MPO are nonnegative, which substantially improves the procedure's numerical stability.

\subsubsection{Truncation}
Whenever a gate acts on a state, the following network update takes place
\begin{equation}
\label{mpsMpo}
\tilde{M}^{s_i}_{(\gamma\bar\gamma)(\rho\bar\rho)} 
=
\sum_{s_i'}  W^{s_i s_i'}_{\gamma\rho} M^{s_i'}_{\bar\gamma\bar\rho}.
\end{equation}
This is depicted in Fig.~\ref{fig:mpo_mps}. Here, a MPO is being absorbed into a MPS at the cost of increasing bond dimension 
(here by a factor of $D=2$). Therefore, a consecutive application of all gates would result in an exponentially large bond dimensions. Hence, the need for a truncation scheme. 
The latter is usually the predominant source of errors~\cite{Murg07, Schollwock_10}. 

Fortunately, such truncation can be managed systematically by looking for a MPS with the smaller (truncated) bond dimension $\chi$. It is found by maximizing its overlap with the original one \cite{Murg07}.  That is to say, one maximizes $|\langle \tilde u | \overline u \rangle|$ between normalized states $| \overline u \rangle$ and $| \tilde u \rangle = F|u\rangle$ as depicted in  Fig.~\ref{fig:mpo_mps}. It is the standard variational approach,  see, e.g., the Ref.~\cite{Schollwock_10}, which we employ in this article.  The general problem of finding optimal MPS matrices $\overline M$ specifying state $| \overline u \rangle$ is highly nonlinear. For that reason, one proceeds site by site,
finding an optimal $\overline M$ for one site while keeping the rest of them fixed. This procedure is repeated while sweeping the chain back and forth until convergence. In practice, this algorithm requires a good starting point to avoid getting trapped in some local extrema.

One could also take advantage of the truncation based on singular value decomposition (SVD).
Therein, the Schmidt decomposition is performed between two parts of the chain, left and right,
\begin{equation}
\label{Schmidt}
|\tilde u \rangle = \sum_k s_k | \tilde u^L_k \rangle | \tilde u^R_k  \rangle.
\end{equation}
The truncation at a given link is then performed by keeping only $\chi$ largest Schmidt values $s_k$, which is optimal from the point of view of a single bond. The error associated with discarding those Schmidt values is $\epsilon=\sqrt{\sum_{k={\chi+1,\ldots}} s^2_k}$. The truncation is performed sequentially on all bonds. In this article, we use the SVD based truncation scheme as an initial condition for the variational procedure. The overlap (fidelity) between the original MPS and the truncated one gives the error associated with the truncation.

Other truncation schemes are also possible. It is worth mentioning that MPO tensors defined in Eqs.~\eqref{gmpo} and \eqref{BC_definition} would render the MPS tensors $M$ nonnegative (assuming no truncation or canonization). This feature may be desirable, both theoretically and numerically, when working with the probability distribution for a classical system~\cite{Temme_stochastic_MPS}.
We should note, however, that the truncation procedure outlined above does not preserve this property. The negative numbers do appear, e.g., in the vectors spanning Schmidt basis. Alternatively, one could use decomposition based on nonnegative matrix factorization. Such an idea was explored, e.g., in the context of simulations of nonequilibrium 1D classical systems with MPS~\cite{Jaksch_classical_MPS}. Nevertheless, the results of that work suggest that SVD based approach provides better numerical accuracy and stability.

As a final note, we would like to stress that it is essential to gradually simulate imaginary time evolution reaching $\beta/2$, by using the gates with smaller $\tau = d\beta$, see Ref.~\cite{cuda_2018} for examples. Even though all gates formally commute, this is not necessarily the case for numerical simulations with finite precision.  For large $\tau$, all gates become ill-conditioned as they approach projectors. That, in practice, may trap the state in Eq.~\eqref{eq:mps} at a local minimum.

\begin{figure} [t]
\begin{center}
  \includegraphics[width=\columnwidth]{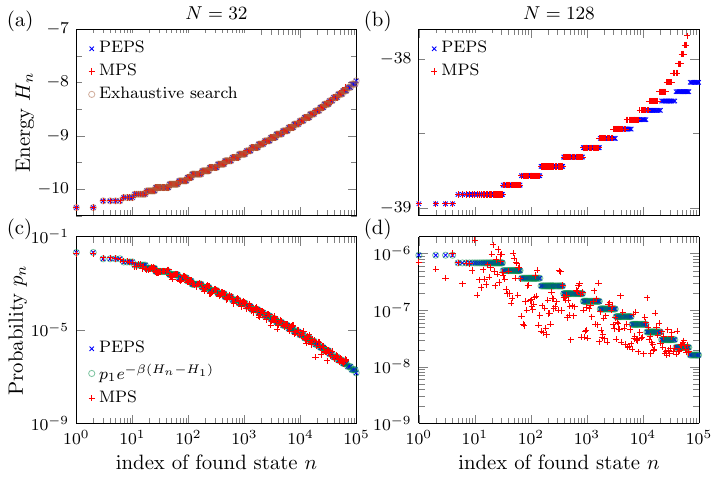}
\end{center}
  \caption{{Comparison between MPS- and PEPS-based approaches.} 
 Top panels show the low-energy spectrum for small chimera graphs with $N=32$ in panel (a), and $N=128$ in panel (b). Bottom panels depict the corresponding probabilities for $N=32$ in panel (c) and $N=128$ in panel (d). While the MPS based approach is still capable of finding the ground state configurations for small systems, the PEPS based approach reveals its superiority. Here, we used $\beta = 5$ and the MPS bond dimension $\chi=128$. All instances are drawn from discreet distribution of $J_{ij}$ with $dJ = \frac{1}{32}$ resulting in discreet values of energies. }
   \label{fig:mps_results}
\end{figure}

\subsubsection{Results}
Having an approximation of the state in Eq.~\eqref{eq:mps} enables one to calculate any conditional probability. Indeed, 
$p{(s_1, s_2, \ldots, s_k)} \approx \langle \rho| \mathcal{P}_{(s_1, s_2, \ldots, s_k)} |\rho\rangle$, where $ | \rho \rangle$ is normalized and $ \mathcal{P}_{(s_1, s_2, \ldots, s_k)} $ is an operator projecting on a given configuration. Calculations of expectation values (or sampling \cite{Ferris12}) of a given MPS can be executed efficiently and exactly \cite{Schollwock_10}. Therefore, after preparing the state $ | \rho \rangle$, we can execute the branch and bound search strategy introduced in the main text.

The results for chimera graphs of sizes $N=32$ and $N=128$ are shown in Fig.~\ref{fig:mps_results}, where we compare them with the PEPS-based approach of the main text. For a very small system size, $N=32$, it is also possible to make a comparison to exhaustive search (brute-force) of low-energy states. In this case, the three are in perfect agreement with each other.  For $N=128$ the MPS-based approach is still able to localize a large set of low-energy states, yet not all of them. As the system size grows, the 1D ansatz loses the capability to capture the physics of the quasi-2D structure faithfully. It is visible in the disparity between the probabilities calculated with MPS and PEPS-based approaches. The latter overlap very well with the Boltzmann factors calculated from energies. The PEPS-based approach can satisfy such a self-consistency check also for large system sizes ($N\sim10^3$), as shown in Fig.~\ref{fig:results}(c) of the main text. The above results provide a perfect setting to appreciate the performance of a PEPS-based approach from the main text in the case of quasi-2D  (chimera in the presented case) graphs. 

Nevertheless, the MPS-based approach discussed above is not limited, at least at the construction level, to the graph's specific geometry. 
It is natural to expect that it would excel for a quasi-1D structure, still allowing for occasional interactions across the chain spanning the problem. We further test the excellent performance of such an approach against exact brute-force search for random fully-connected problems up to $N=50$ in Ref.~\cite{cuda_2018}.

\subsection{Efficient calculation of probabilities}
\begin{figure} [t]
    \begin{center}
        \includegraphics[width=\columnwidth]{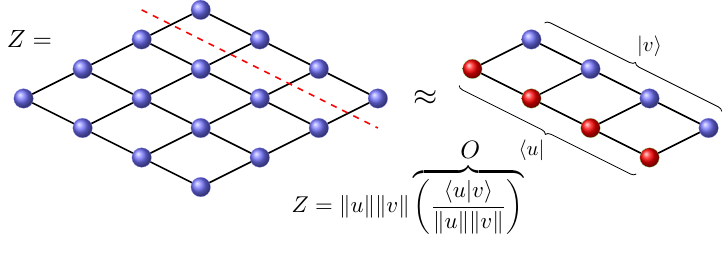}
    \end{center}
    \caption{{Calculation of the partition function using boundary MPSs}.  
        The overlap $O$ between normalized MPSs marked as $|u\rangle$ and  $|v\rangle$ reflects on numerical stability of the problem.
    }
    \label{fig:overlap}
\end{figure}

A contraction of the PEPS tensor network is necessary to extract information regarding the marginal and conditional probabilities.
In this article, we use a boundary-MPS based approach for that purpose \cite{Murg07, Cirac14}. In particular, the techniques briefly described above in the context of the MPS-based algorithm can be directly applied here.
Indeed, after tracing (or projecting) out physical degrees of freedom, a PEPS tensor [see Eq.~\eqref{cpeps} in the main text] can be reinterpreted as an MPO, i.e.,
\begin{equation}
\label{update}
A^{\bfm{ud}}_{\bfm{lr}} 
=
\sum_{\bfm s_{i}} A^{\bfm s_i}_{\bfm{lrud}}. 
\end{equation}
Therefore, the MPO-MPS contraction scheme~\eqref{mpsMpo} can be applied to collapse the PEPS tensor network, layer by layer, starting from the bottom up (or from the top down, etc.). This is exactly how we proceed in this article; see the transition between Figs.~\ref{fig:all}(e) and~\ref{fig:all}(f). 
As a preprocessing step, we begin with the initial preparation of boundary MPSs representing two, three, etc. rows of PEPS tensors. One such boundary MPS, corresponding to two rows, is marked as green
in Fig.~\ref{fig:all}(e). For instance, to obtain the partition function, one then calculates the overlap of the boundary MPSs representing respectively the top and bottom part of the network, as depicted in Fig.~\ref{fig:overlap}. 

The leading numerical cost is related to the truncation of the boundary MPS. In the approach we employ, it scales as $O(N D^4 \chi^3)$. That is, the leading cost of obtaining the Schmidt decomposition in Eq.~\eqref{Schmidt} for the enlarged MPS tensors of size $\chi D \times D \times \chi D$. To that end, for each site, one needs to calculate the QR (or SVD) decomposition of  $\chi D^2 \times \chi D$ matrix at a cost $O(\chi^3 D^4)$. A less accurate initial guess for a subsequent variational optimization may be found at a lower numerical cost \cite{Schollwock_10}. The tensor contractions needed for variational optimization are similar to calculating the MPS-MPO-MPS expectation value. In our case this is executed at a cost $O[N (D^2 \chi^3 + D^4 \chi^2)]$. Finally, tracing out the spin degrees of freedom of PEPS tensors is done at a cost $O(N D^4 d^l)$.

Subsequently, to calculate the marginal conditional probabilities, 
\begin{equation}
\label{eq:cond}
\includegraphics[width=0.9\columnwidth]{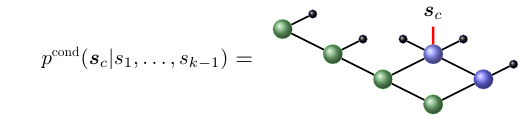}
\end{equation}
one focuses on a given configuration, $s_{1}, \ldots, s_{k-1}$, spanning the upper half of the lattice. 
Above, the black dots represent tensors $B$ or $C$ completing the decomposition in Eq.~\eqref{asvd}, projected on this configuration.  This procedure allows one to calculate all the probabilities invoked while executing the branch and bound strategy from the main text---where we explore the lattice row after row. The leading numerical cost of contracting such a network is 
$O[N M  (\chi^2 D^2 + D^2 d^l)]$, assuming here that \textit{$\chi > D$}. 
Note that partial contractions can be cached for efficiency when calculating probabilities of consecutive sites along a row.

\subsection{Conditioning and compact representation}
\label{sec:cond}
\begin{figure} [t]
\begin{center}
  \includegraphics[width=\columnwidth]{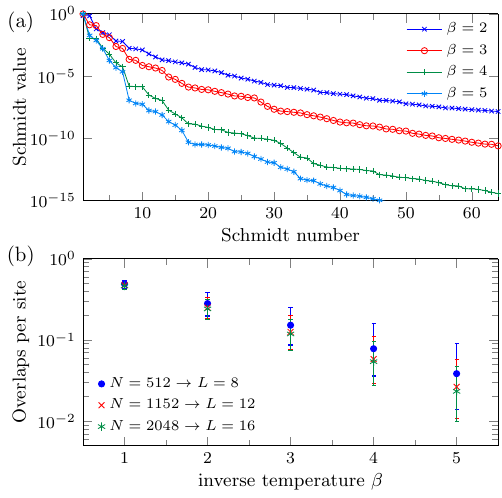}
\end{center}
  \caption{{Schmidt values and overlaps of boundary MPS}. In panel (a), we show the decay of the Schmidt values for boundary MPS representing all but the last layer of the network, cf. Fig.~\ref{fig:overlap}, split in the middle. Results for chimera graph with $N=2048$, corresponding to Fig.~\ref{fig:results} of the main text. The spectrum is quickly decaying with growing $\beta$.  (b) The respective overlap per site $O^{1/L}$, cf. Fig.~\ref{fig:overlap}. It is decaying with growing $\beta$ indicating ill-conditioning of the problem.
The collapse of curves for different linear system sizes $L$ points out that $O$ is vanishing exponentially with $L$. The plot shows a median of 100 {\it droplet instances} with the error bars corresponding to 1-sigma of the distribution. $L$ is defined here as the length of boundary MPS used to contract the network. 
Results were obtained after employing the preconditioning procedure outlined in the text.
  }
   \label{fig:schmidt}
\end{figure}

The feasibility of the outlined approach hinges heavily on the existence of a faithful representation of boundary MPSs with a small enough bound dimension.
The latter can be assured by quickly decaying Schmidt spectrum; see Eq.~\eqref{Schmidt}.  When only $\chi$ largest Schmidt values are kept, the error can be quantified with discarded Schmidt values. A typical Schmidt spectrum, shown in Fig.~\ref{fig:schmidt}(a), was calculated in the middle of boundary MPS. The latter captures all but the last layer of PEPS for a single instance (corresponding to Fig.~\ref{fig:all} in the main text). 
As is evident, the Schmidt spectrum is vanishing rapidly, indicating that a compact representation indeed exists.
Importantly,  increasing the value of $\beta$ causes the Schmidt spectrum to vanish more rapidly -- the point which we are going to elaborate on a bit more shortly.

However, the partition function in Fig.~\ref{fig:overlap}, or, more importantly, probabilities in Eq.~\eqref{eq:cond}, are effectively calculated as an overlap between two MPSs that
represent lower and upper parts of the network.  For the sake of clarity, we focus on the overlap between two normalized vectors (MPSs), shown in Fig.~\ref{fig:overlap}.  The boundary MPS $|u \rangle$ approximates the exact one, $|u \rangle + |\epsilon_u \rangle$,  with an error  $\epsilon_u = || \epsilon_u||_2$ given by 2-norm. Hence, the overlap error can be bounded by $\epsilon_O = |\langle \epsilon_u| v \rangle| \le \epsilon_u$.
It illustrates that when the overlap $O$ is decreasing, one would desire $\epsilon_u$ to be sufficiently smaller to maintain the relative error under control.
As a result, the overlap $O$ provides a direct indication of the problem's conditioning.
Note that this discussion directly extends to unnormalized marginal probabilities in Eq.~\eqref{eq:cond}, which sum up to $O$ (perhaps calculated for a subsystem). 

\begin{figure} [t]
\begin{center}
  \includegraphics[width=0.90\columnwidth]{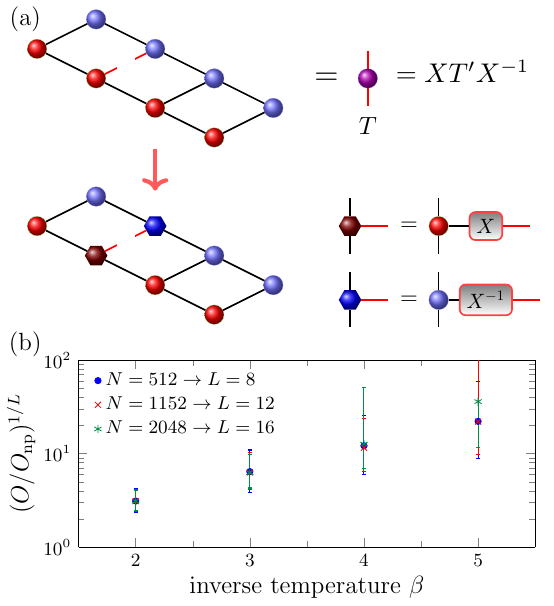}
\end{center}
  \caption{ {Preconditioning of PEPS tensors.} (a) The overlap between two boundary MPSs, resulting in the partition function, is calculated up to contraction of a single bound. This gives matrix $T$. We use balancing of $T$ to find gauge transformation for the PEPS network. 
  We observe that it often leads to increased overlap. This is shown in panel (b) where we compare the overlap per site with preconditioning, $O$, and with no preconditioning, $O_{\rm {np}}$. We present median of 100 instances shown in Fig.~\ref{fig:schmidt}. }
   \label{fig:mp}
\end{figure}

One can naturally expect $O$ to vanish exponentially with the linear system size $L$, as an overlap of two vectors in large space of dimension $\sim {D^L}$. Indeed,  in Fig.~\ref{fig:schmidt}(b) we show a typical overlap per site, $O^{1/L}$,  as a function of $\beta$. The data for different $N$ (which translates to $L$), obtained for {\it droplet instances}, indeed coincide. Moreover, the calculated points vanish quickly with $\beta$, indicating a possible need for greater accuracy. 

It clearly illustrates the trade-off when choosing the control parameters for the algorithm. On the one hand, larger $\beta$ are preferable, as they allow one to ``zoom in" on the low-energy spectrum.
On the other hand, this inevitably leads to problem conditioning. Indeed, if too large $\beta$ is used, then the probabilities cannot be calculated with the desired precision. While the efficient boundary MPSs exist, it may require increasing numerical precision to capture sufficiently small Schmidt values -- similarly as was observed for simulation of stochastic processes using MPS \cite{Jaksch_classical_MPS, Carlon1999}.
Nevertheless, the standard 64-bit numerical precision used in this work seems to be enough to emulate the problem sizes available on the current quantum annealers (at least for problems classes considered here).

In practice, one should start with small enough values of $\beta$. Nevertheless, what is small may depend on a particular instance or instance set if they are not random but generated according to some heuristics. Subsequently, $\beta$ can be increased as long as it allows to obtain self-consistent results.

\subsection{Preconditioning of boundary MPS}
\label{precond}
We can use the insight from the previous section to set up a preconditioning procedure for the PEPS network. Its tensors are defined up to a local gauge transformation, which reflects on nonuniqueness of the decomposition in Eq.~\eqref{asvd} or Eq.~\eqref{BC_definition}. The idea is to insert a resolution of identity, $X X^{-1}$, on each virtual bound to increase the overlap between the boundary MPSs, cf. Fig.~\ref{fig:mp}.

In principle, finding appropriate $X$'s and $X^{-1}$'s is hard. Moreover, it is easy to introduce numerical instabilities with careless choices. For that reason, we limit ourselves to diagonal $X$'s with positive elements on the diagonal, with inverse having the same properties. Consequently, after applying gauge tensors, the PEPS tensors remain composed of nonnegative elements, which typically is a good choice to retain the numerical stability.

To find the preconditioning, we focus on one link at a time and proceed as follows. We contract all the other links forming an overlap, as depicted in Fig.~\ref{fig:mp}. The remaining object, marked as $T$ in that figure, can be regarded as a matrix, the trace of which gives the overlap. However, the off-diagonal elements of $T$ are usually large in comparison to the diagonal ones, which also reflect on the conditioning of the contraction.

We observe that good results are often obtained by applying heuristic procedure based on a balancing scaling transformation \cite{balancing}, $T = X T' X^{-1}$, as usually implemented in numerical libraries. The aim of this procedure is to balance the $1$-norm of rows and columns of the matrix. It is a standard preconditioning procedure invoked when numerically finding eigenvalues and eigenvectors. The far-fetched idea is that in the ideal case when the overlap is $1$, $T$ would be symmetric. Such $X$ has the desired property of being diagonal and positively defined, which preserves the nonnegativity of PEPS tensors obtained with the building blocks in Eq.~\eqref{BC_definition}. Other possible strategies to find gauges exist, e.g., trying to directly maximize the overlap $O$, which we do not explore here.

We find the scaling transformations for a smaller value of $\beta$, for which the overlap and conditioning are better. These gauge transformations are then applied to the virtual indices of PEPS tensors for larger, target $\beta$, for which all the probabilities are calculated. We typically employ preconditioning procedures at $\beta/4$ and $\beta/2$ and find that this often increases the method's stability. The added numerical cost is the same as for calculating the preprocessing step where the boundary MPSs are found. We show the overlaps resulting from such preconditioning for {\it droplet instances} in Fig.~\ref{fig:schmidt}(b). We note, however, that different heuristic preconditioning procedures might prove effective for different classes of instances, which we leave for a later contribution.

\subsection{Conditioning of MPO's}
\begin{figure} [b]
\begin{center}
  \includegraphics[width=0.9\columnwidth]{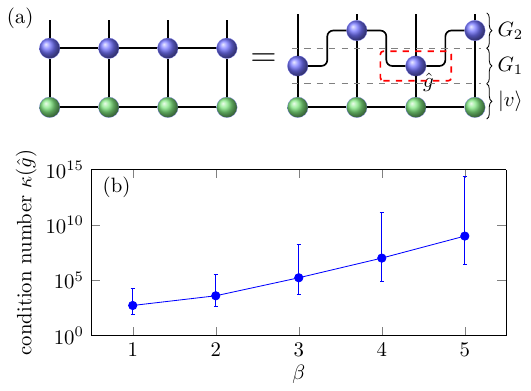}
\end{center}
  \caption{{Conditioning of one layer of MPO}. (a) MPO (blue) can be viewed as a series of local gates $\hat g$ acting on a boundary MPS (green). In panel (b), we show condition number of $\hat g$---viewed as a matrix. It is growing quickly with increasing $\beta$, indicating that larger $\beta$ should make contraction of PEPS network less reliable.  We plot the  median value with the error bars indicating 1-sigma of the distribution. Data for {\it droplet instances}.}
   \label{fig:conditioning}
\end{figure}

The error related to a single truncation of boundary MPS is well controlled and quantified by the overlap between MPS before and after truncation, as discussed in previous sections. We should note, however, that the PEPS network itself is typically ill-conditioned. As such, the relative error resulting from previous truncations, or finite numerical precision, can be effectively amplified (or reduced) by the application of consecutive layers of MPO. This is depicted in Fig.~\ref{fig:conditioning}(a), where the green boundary MPS $| u \rangle$ is an approximation of the exact one $| \tilde u\rangle + |\epsilon_u \rangle$. Acting with an MPO on that MPS can be viewed as a series of {\it local} gates. They can be divided, e.g., into $G_1$ and $G_2$ as depicted in the figure.

We plot the condition number of such a single gate, marked as $\hat g$ and treated as a matrix, in Fig.~\ref{fig:conditioning}(b). The condition number,  i.e., the ratio between the largest and smallest singular values of the matrix, gives a bound on how much the relative error may grow in the worst-case scenario.  As can be seen in the figure, the condition number is growing quickly with increasing $\beta$, which is in agreement with our previous argument that the large $\beta$ renders contraction of the network more difficult.

Consecutive application of local gates may, in the worst-case, result in an error growing exponentially with the system size.
Nevertheless, all the evidence from extensive numerical simulations suggests that such worst-case is often not happening in practice. It is in accordance with a general observation that truncation of the PEPS tensor network usually can produce reliable results beyond what is suggested by the worst-case bounds, see, e.g., Ref.~\cite{Haegeman16, gluza_peps_2018}.

Furthermore, we can speculate that a better understanding of truncation errors, their potential locality, and their relation with the frustration of the problem could allow one to obtain much tighter bounds on the error propagation. Note that if the errors are local (along the boundary MPS), then the worst-case bounds related to local gates $\hat g$  would add up and not multiply.  This could then formally help to certify the solution, at least for sufficiently small problems.
\end{appendix}


%\bibliographystyle{apsrev4-1}
%\bibliography{bib.bib} 

%======================================================================
 \end{document}